\documentclass[%
 reprint,
 amsmath,amssymb,
 aps,
floatfix,
]{revtex4-2}
\usepackage{graphicx}% Include figure files
\usepackage{dcolumn}% Align table columns on decimal point
\usepackage{bm}% bold math
\usepackage{hyperref}% add hypertext capabilities
\usepackage{cleveref}
\usepackage[mathlines]{lineno}% Enable numbering of text and display math
\usepackage{draftwatermark}
\SetWatermarkText{DRAFT}
\SetWatermarkScale{5}
\begin{document}

\preprint{APS/123-QED}

% \title{Manuscript Title:\\with Forced Linebreak}% Force line breaks with \\
% \title{The Resilience of Propagating Compaction Bands in Brittle Porous Media}
% \title{ Propagating Compaction Bands Near Shaped Boundaries in Brittle Porous Media}
\title{A consistent derivation of soil stiffness from elastic wave speeds}

\author{David M. Riley}
\author{Itai Einav}
\author{Fran\c{c}ois Guillard}
\affiliation{Particles and Grains Laboratory, School of Civil Engineering, The University of Sydney,
Sydney, New South Wales 2006, Australia}

% \author{Julio Valdes}%
%  \email{Second.Author@institution.edu}
% \affiliation{%
%  Geo-Innovations Research Laboratory, Department of Civil, Construction, and Environmental Engineering,
% San Diego State University, San Diego, California 92182, USA}%

% \collaboration{MUSO Collaboration}%\noaffiliation

%  \homepage{http://www.Second.institution.edu/~Charlie.Author}
% \affiliation{
%  Second institution and/or address\\
%  This line break forced% with \\
% }%
% \affiliation{
%  Third institution, the second for Charlie Author
% \affiliation{%
%  Authors' institution and/or address\\
%  This line break forced with \textbackslash\textbackslash
% }%

% \collaboration{CLEO Collaboration}%\noaffiliation

\date{\today}% It is always \today, today,

\begin{abstract}
Elastic wave speeds are fundamental in geomechanics and have historically been described by an analytic formula that assumes \textit{linearly} elastic solid medium. Empirical relations stemming from this assumption were used to determine \textit{nonlinearly} elastic stiffness relations that depend on pressure, density, and other state variables. Evidently, this approach introduces a mathematical and physical disconnect between the derivation of the analytical wave speed (and thus stiffness) and the empirically generated stiffness constants. In our study, we derive wave speeds for energy-conserving (hyperelastic) and non-energy-conserving (hypoelastic) constitutive models that have a general dependence on pressure and density. Under isotropic compression states, the analytical solutions for both models converge to previously documented empirical relations. 
Conversely, in the presence of shear, hyperelasticity predicts changes in the longitudinal and transverse wave speed ratio. This prediction arises from terms that ensure energy conservation in the hyperelastic model, without needing fabric to predict such an evolution, as was sometimes assumed in previous investigations. Such insights from hyperelasticity could explain the previously unaccounted-for evolution of longitudinal wave speeds in oedometric compression.  Finally, the procedure used herein is general and could be extended to account for other relevant state variables of soils, such as grain-size, grain-shape, or saturation.

% \begin{description}
% Secondary publications and information retrieval purposes.
% You may use the \texttt{description} environment to structure your abstract;
% use the optional argument of the \verb+\item+ command to give the category of each item. 
% \end{description}
\end{abstract}

\maketitle

\section{Introduction}
Accurate determination of wave speeds and small-strain stiffness is crucial for a thorough geotechnical evaluation of soil sediments~\citep{atkinson2000non,mayne2001stress} and the precise design of geotechnical structures~\citep{clayton2011stiffness}. In geomechanics, the wave speeds are conventionally assumed to be equivalent to the ones derived for linear elastic materials~\cite{hardin1963elastic,hardin1989elasticity,wichtmann2009influence,li2022evaluation}. Based on this assumption, empirical relations are obtained for the nonlinear stiffness and are found to be dependent on pressure, density, and other state variables. However, assuming the wave speeds are defined for a linear elastic solid is generally an oversimplification for porous media and inconsistent with empirical relationships, necessitating the exploration of more general nonlinear elastic models. Here, wave speeds are derived for non-energy-conserving hypoelastic as well as energy-conserving hyperelastic models, assuming a general stiffness dependence on pressure and density while ensuring mathematical consistency. 
% The comparison of these two models also allows for exploring the impact of energy conservative models (hyperelasticity).

Conventionally, in geotechnical engineering, wave speeds are assumed to be characterised by a homogeneous, isotropic, linearly elastic solid. The resulting longitudinal and transverse wave speeds are defined as:
\begin{equation} \label{eq: vp_elastic}
    V_p = \sqrt{\frac{M_w}{\rho}},
\end{equation}
\begin{equation} \label{eq: vs_elastic}
    V_s = \sqrt{\frac{G_w}{\rho}},
\end{equation}
where $M_w$ and $G_w$ are the constrained and shear moduli, respectively. After experimentally measuring the wave speeds, the stiffness moduli $M_w$ and $G_w$ are found to depend on the effective pressure $p$ and solid fraction $\phi$ in soils. Generally, these empirical relations adopt the form~\citep{hardin1963elastic}:
\begin{equation} \label{eq: M_iso_emp}
    M_w = A H(\phi) \left( \frac{p}{p_a}\right)^{b} ,
\end{equation}
\begin{equation} \label{eq: G_iso_emp}
    G_w = B H(\phi) \left( \frac{p}{p_a}\right)^{b} ,
\end{equation}
where $A$ and $B$ are constants, $p_a= 1$ Pa is a reference stress, $b$ is the stress exponent reflecting nonlinear pressure dependency, and $H(\phi)$ is a general function of solid fraction $\phi$. Note that the void ratio $e=\frac{1-\phi}{\phi}$ is normally adopted in geotechnical engineering rather than the solid fraction $\phi$, but we opt for a description in terms of solid fraction for convenience. Typically, $\frac{1}{3} \lesssim b \lesssim \frac{1}{2}$ for particles spanning from perfectly spherical to angular~\citep{santamarina1998effect}, as validated through effective medium representation~\citep{yimsiri2000micromechanics,agnolin2007internal,goddard1990nonlinear}. Furthermore, more advanced relations have been proposed to include general stress states~\citep{roesler1979anisotropic,yu1984stress,hardin1989elasticity,bellotti1996anisotropy} and other state variables such as grain-size distribution~\citep{wichtmann2009influence,wichtmann2010influence,payan2017effect}, particle shape~\citep{liu2018shear,tang2021wave}, or fabric~\citep{mital2020effect,li2022evaluation}. However, within the current paper, we do not explore these options as we seek to solely understand the effect of pressure and density dependence on the wave speed. 

\begin{figure}[h]
	\centering
	\includegraphics[width=\linewidth]{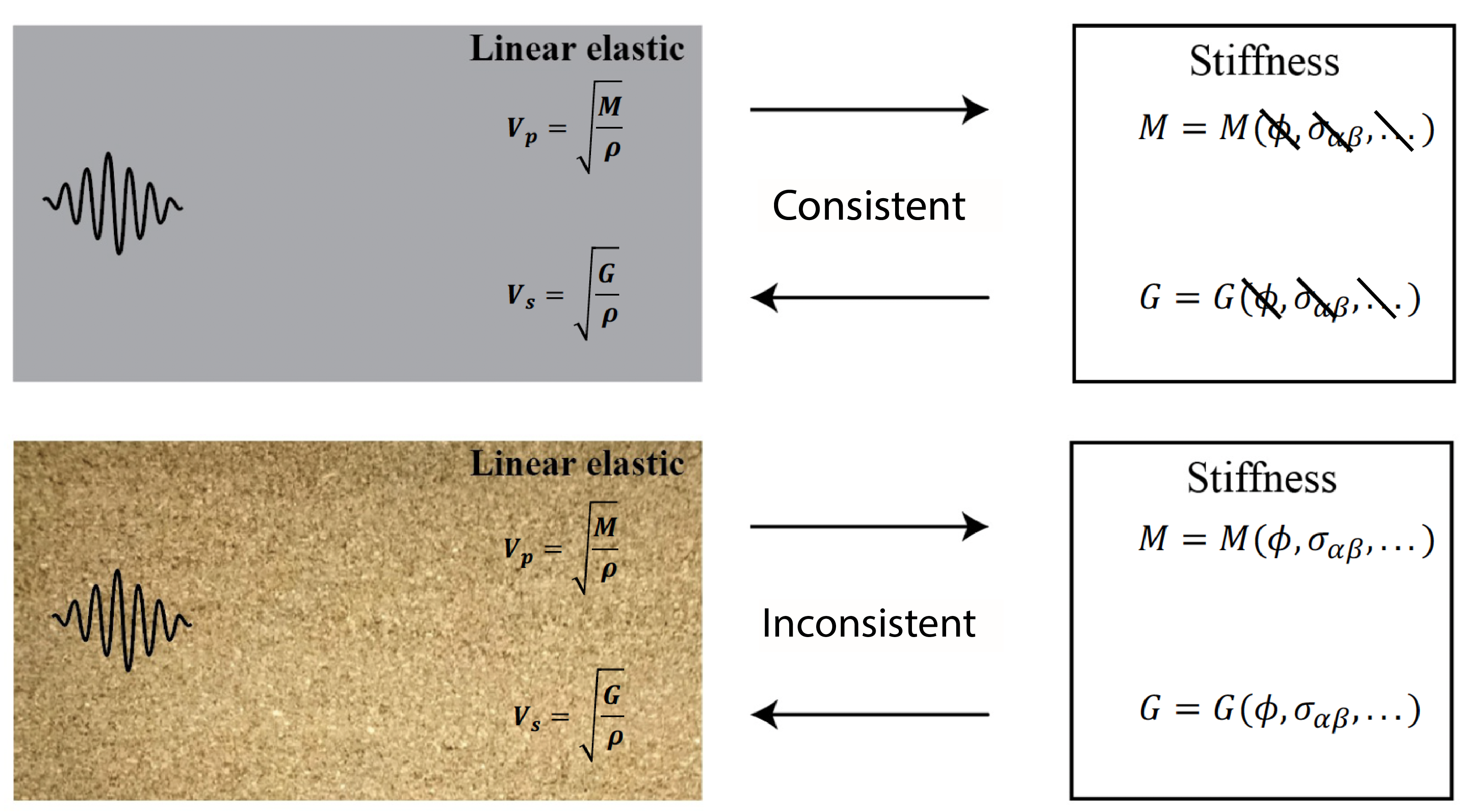}
  	\caption{Illustration of the inconsistency of assuming elastic wave speeds in soils. {\textit{Top row.}} Linear elastic wave speeds are valid for materials in which stiffness is not a function of state. {\textit{Bottom row.}} In soils, linear elastic wave speeds are assumed despite stiffnesses being a function of state. Here $\phi$ is solid fraction and $\sigma_{\alpha\beta} $ is the stress tensor.}
	\label{fig: 0c5}       % Give a unique label
\end{figure}

Although these empirical relations have successfully captured experimental phenomena, it is crucial to note that they were determined by first assuming a linearly elastic isotropic continuum. This introduces a discrepancy between the assumed elastic wave speeds, as given by~\cref{eq: vp_elastic,eq: vs_elastic} and the retrieved empirical equations~\cref{eq: M_iso_emp,eq: G_iso_emp}.~\cref{fig: 0c5} highlights this inconsistency, where soils are known to have state-dependent stiffness, yet state-independent linear elasticity is assumed to derive the wave speeds. Despite existing models that incorporate pressure dependence in their constitutive laws~\citep{jiang2003granular,einav2004pressure,houlsby2005elastic,nguyen2009energetics}, wave speeds are usually not derived with consideration for pressure dependence. However, hyperelastic pressure-dependent models assuming specific pressure dependencies have recovered empirical relations under isotropic compression states~\citep{andreotti2013granular} and uniaxial compression states that showed shearing alters the evolution of the longitudinal to transverse wave speed ratio~\citep{mayer2010propagation}.

% However, analyses using pressure-dependent models during only isotropic compression states have mirrored the pressure dependencies seen in empirical relations~\citep{andreotti2013granular}. Another study used a hyperelastic model with $b=1/3$ for~\cref{eq: M_iso_emp,eq: G_iso_emp} and a density dependence equivalent to the empirical one established in~\cite{hardin1963elastic}, and derived the wave speed .

% Moreover, a study has derived wave speeds for uniaxial compression state of hyperelastic model such that it recovers $b=1/3$ for~\cref{eq: M_iso_emp,eq: G_iso_emp} and hinted that shearing induces changes to the evolution of the ratio of the longitudinal and the transverse wave speeds~\citep{mayer2010propagation}.

% pressure and density-dependent media but assumes coefficients to generate Hertzian pressure scaling~\citep{mayer2010propagation}.

% Yet, such studies of wave speeds in pressure-dependent media often ignore the density dependence of stiffness and, more broadly, the role of shear stresses on these moduli. 

Herein, analytical wave speeds are derived for both hyperelastic and hypoelastic constitutive models with a general pressure and density dependence. Importantly, the passage of energy is conserved in hyperelastic model, but this is not the case for the hypoelastic model. Furthermore, this derivation procedure remains accurate and general, irrespective of the model selected. The derived wave speeds reproduce the empirical relations under isotropic stress conditions regardless of whether hyperelasticity or hypoelasticity is considered. However, due to the density dependence, a minor, typically negligible, correctional term arises for the hyperelastic model. Conversely, when the material undergoes shear, the wave speed derived for hyperelasticity is significantly different than the hypoelastic wave speed because of the presence of additional non-negligible stress-dependent terms. These insights shed light on the evolution of the ratio between longitudinal and transverse wave speeds observed experimentally that previously could not be explained with empirical relations when assuming a linearly elastic isotropic continuum.

% This directly results from two main factors: the hyperelastic model is derived from a potential that ensures energy conservation, and the wave speeds are derived from a consistent mathematical foundation. These insights also shed light on the evolution of the ratio between longitudinal and transverse wave speeds observed experimentally that previously could not be satisfactorily explained with empirical relations when assuming a linearly elastic isotropic continuum.

\section{Wave speed derivation} \label{sect: wave perturb}
We now derive the analytical expressions for longitudinal and transverse wave speeds. The momentum balance is given by
\begin{equation} \label{eq: mom_balance}
    \rho \ddot{u}_\alpha=\sigma_{\alpha\beta,\beta},
\end{equation}
where $\rho$ is the bulk density, $\ddot{u}_\alpha$ represents acceleration in the $\alpha$ direction, $\sigma_{\alpha\beta}$ is the total stress tensor, and $\Box_{,\alpha}=\frac{\partial \Box}{\partial x_{\alpha}}$. For a homogeneously deformed solid at equilibrium, this simplifies to
\begin{equation} \label{eq: equilibrium}
    \sigma_{\alpha\beta,\beta}^{*}=0,
\end{equation}
where $\sigma_{\alpha\beta}^{*}$ is the homogeneous reference (or unperturbed) stress. By first-order Taylor series expansion, the stress can be expressed in terms of $\sigma_{\alpha\beta}$ and a stress perturbation $\tilde{\sigma}_{\alpha\beta}$, and is given by
\begin{equation} \label{eq: stress_peturb}
    \sigma_{\alpha\beta}=\sigma_{\alpha\beta}^*+ \tilde{\sigma}_{\alpha\beta} =\sigma_{\alpha\beta}^{*}+ C_{\alpha\beta\gamma\zeta}^{*} \tilde{u}_{\gamma,\zeta}
\end{equation}
where $C_{\alpha\beta\gamma\zeta}^*$ is the isotropic homogeneous elastic stiffness tensor at the reference state, $\tilde{u}_\alpha$ is a displacement perturbation from the reference homogeneous solution of the displacement field $u_\alpha^*$, such that $u_\alpha = u_\alpha^* + \tilde{u}_\alpha$~\citep{stefanou2016fundamentals,stefanou2019strain}. The strain tensor is defined as $ u_{\gamma,\zeta}=\varepsilon_{\gamma\zeta}$, and thus, $\tilde{u}_{\gamma,\zeta}=\tilde{\varepsilon}_{\gamma\zeta}$ is the perturbed strain tensor. By inserting~\cref{eq: stress_peturb} into~\cref{eq: equilibrium} and using a first-order Taylor expansion of the left-hand side, assuming no acceleration in the reference state ($ {\ddot{u}}_\alpha^* =0$) the momentum balance is expressed as
\begin{equation} \label{eq: perturbed_mom}
    \rho^{*} \tilde{\ddot u}_\alpha = \tilde{\sigma}_{\alpha\beta,\beta}
\end{equation}
where $\rho^{*}$ is the homogeneous equilibrium density. The perturbed stress tensor $\tilde{\sigma}_{\alpha\beta}$ will be later determined from the constitutive law through first-order Taylor series expansion. 

\begin{figure}[!htbp]
	\centering
	\includegraphics[width=\linewidth]{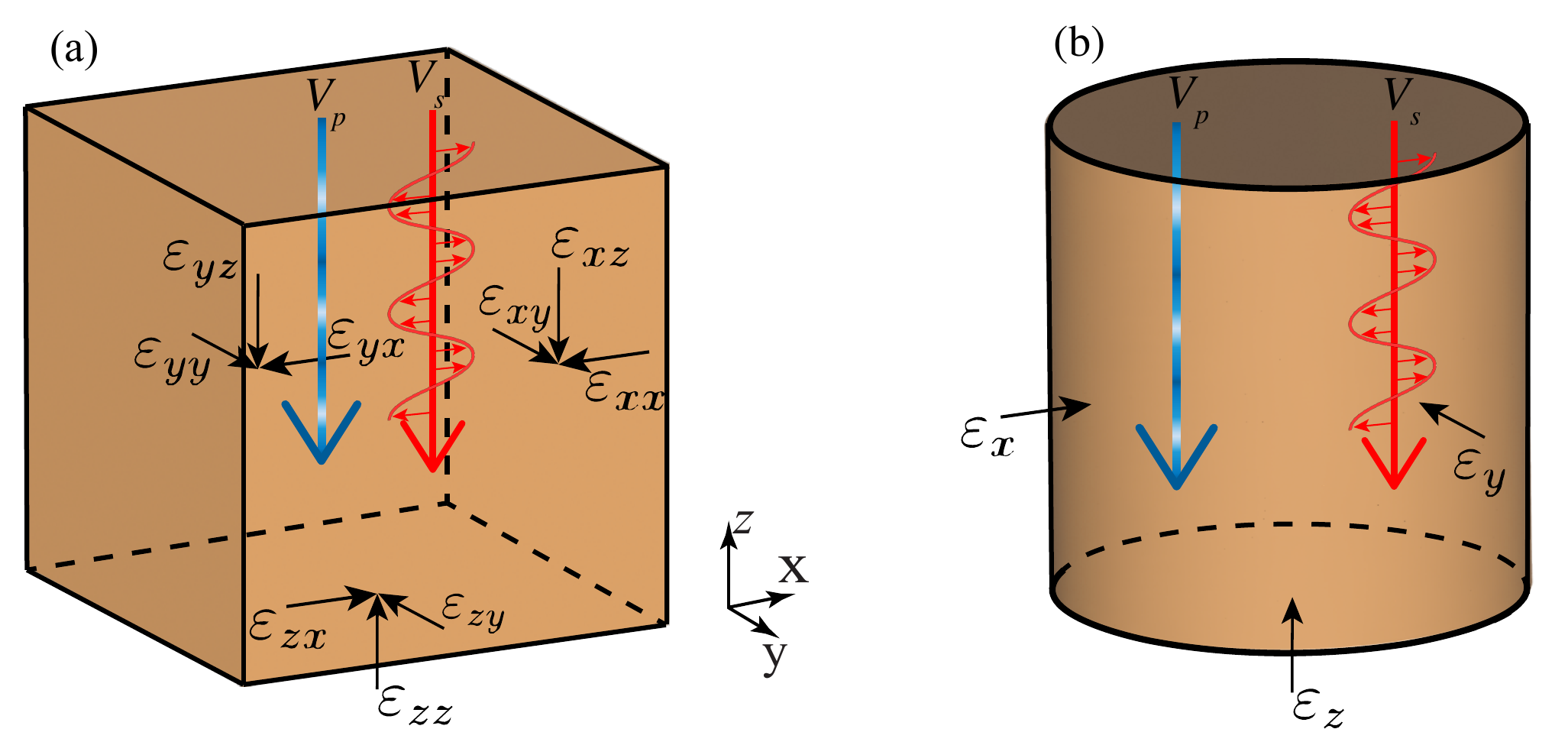}
 \caption{Illustration of longitudinal wave (blue gradient line) and transverse wave (red) for (a) general case and (b) in triaxial conditions. For the longitudinal wave, the periodic gradient denotes material displaced in the z-direction, and the transverse wave has the material displaced in the x-direction, denoted by small red arrows. Both waves propagate in the z-direction. }
  	% \caption{Illustration of longitudinal wave (blue gradient line) and transverse wave (red) for (a) general case and (b) in triaxial conditions used in~\cref{sect: triax}. For the longitudinal wave, the periodic gradient denotes particles displaced in the z-direction, and the transverse wave has the material displacing in the x-direction, denoted by small red arrows. Both waves propagate in the z-direction. }
	\label{fig: 1c5}       % Give a unique label
\end{figure}

First, consider longitudinal waves that propagate along the z-axis. The perturbed displacement $\tilde{u}_z$ is decomposed in Fourier modes of the form $\tilde{u}_z=U_z e^{i(kz-\omega t)}$, where $i$ is the unit imaginary number, $k$ is the wave number, and $\omega$ is the circular frequency. The perturbed strain tensor and its deviatoric component are thus expressed as
\begin{equation} \label{eq: long_pertub}
    \tilde{\varepsilon}_{\alpha\beta}=
ik\Tilde{u}_z\begin{bmatrix}
0 & 0 & 0\\
0 & 0 & 0\\
0 & 0 & 1
\end{bmatrix}, \quad \tilde{\varepsilon}_{\alpha\beta}^{\prime}=
ik\Tilde{u}_z\begin{bmatrix}
-\frac{1}{3} & 0 & 0\\
0 & -\frac{1}{3} & 0\\
0 & 0 & \frac{2}{3}
\end{bmatrix},
\end{equation}
where $\tilde{\varepsilon}_{\alpha\beta}^{\prime} = \tilde{\varepsilon}_{\alpha\beta}-\frac{1}{3}\tilde{\varepsilon}_{\gamma\gamma}\delta_{\alpha\beta}$ is the deviatoric component of the perturbed strain tensor with $\delta_{\alpha\beta}$ being the Kronecker delta.

For a transverse wave, we assume that the wave propagates in the z-axis and is polarised in the x-direction, where the perturbed displacement  $\tilde{u}_x=U_x e^{i(kz-\omega t)}$. The disturbance to the strain field and its deviatoric component for the transverse wave are equal in this case and given by
\begin{equation} \label{eq: shear_perturb}
    \tilde{\varepsilon}_{\alpha\beta}=\tilde{\varepsilon}_{\alpha\beta}^{\prime}=
ik\Tilde{u}_x\begin{bmatrix}
0 & 0 & \frac{1}{2}\\
0  & 0 & 0\\
\frac{1}{2} & 0 & 0
\end{bmatrix}.
\end{equation}

~\cref{fig: 1c5} (a) illustrates these wave perturbations for a general state, where the periodic blue arrow shows the longitudinal wave and the red displays the transverse wave. The periodic gradient shows that the particle motion is in the same direction as the wave propagation for the longitudinal wave. In contrast, the particle motion occurs perpendicular to the wave direction in the transverse wave case.

To determine the wave speed, a constitutive model is required to define the isotropic homogeneous elastic stiffness tensor, as this enables an explicit representation of~\cref{eq: perturbed_mom} following the substitution of the perturbed strains into the perturbed stress. The following section describes two potential constitutive relations to apply these strain perturbations.

\subsection{Hyperelastic wave speed} \label{sect: hyper_wave}
Now that a general framework for wave speed derivations has been laid, a constitutive law is required to evaluate the perturbed stress. We begin by introducing a hyperelastic model for pressure and density-dependent media. The advantage of hyperelasticity is that it ensures the energy is being conserved as waves travel through the medium, thus keeping faithful to the meaning of a true elastic body. Here, we consider an internal energy density $\mathcal{U}$ that broadens previous work by~\cite{alaei2021hydrodynamic}, and depends on solid fraction $\phi$ and elastic strain tensor $\varepsilon_{\alpha\beta}$:
 \begin{equation} \label{eq: internal energy}
     \mathcal{U} = F(\phi)\left(\frac{\bar{K}}{(1+n)(2+n)}\varepsilon_v^{n+2} +\bar{G} \varepsilon_v^{n}\varepsilon_{\alpha\beta}^{\prime}\varepsilon_{\alpha\beta}^{\prime} \right)
 \end{equation}
 where $\bar{K}$ is the bulk stiffness constant, $\bar{G}$ is the shear stiffness constants, respectively; and $\varepsilon_v=\varepsilon_{\alpha\alpha}$ is the volumetric elastic strain. The choice of this energy form will be justified later. 
 %For simplicity, we choose a simple solid fraction dependence of the form $F(\phi)=\phi^{\bar{d}}$, with $\bar{d}$ as a constant, where this form was motivated by experiments~\citep{rubin2011large}. To underscore the generality of the subsequent derivations, we refrain from substituting this function at this stage. 
 %Furthermore, we note that the triaxial shear strain invariant is given by $\varepsilon_s=\sqrt{\frac{2}{3}\varepsilon_{\alpha\beta}^{\prime}\varepsilon_{\alpha\beta}^{\prime}}$. 
 The pressure and deviatoric stress are given through differentiation of the internal energy density
\begin{equation}  \label{eq: hyper_p}
    p=\frac{\partial \mathcal{U}}{\partial \varepsilon_{v}} = F(\phi) \left( \frac{\bar{K}}{n+1}\varepsilon_v^{n+1} + \frac{3}{2}n \bar{G} \varepsilon_{v}^{n-1} \varepsilon_s^2 \right),
\end{equation}
\begin{equation} \label{eq: hyper_dev}
    \sigma_{\alpha \beta}^{\prime}= \frac{\partial \mathcal{U}}{\partial \varepsilon_{\alpha\beta}^{\prime}}=  2F(\phi)\bar{G} \varepsilon_v^n \varepsilon_{\alpha \beta}^{\prime},
\end{equation}
where, $\sigma_{\alpha\beta}=\sigma_{\alpha\beta}^{\prime}+p\delta_{\alpha\beta}$ is the total stress tensor, and $\varepsilon_s=\sqrt{\frac{2}{3}\varepsilon_{\alpha\beta}^{\prime}\varepsilon_{\alpha\beta}^{\prime}}$ is the triaxial shear strain invariant.
It is worth noting the generality of the model above. Typically, we will choose $F(\phi)=\phi^{\bar{d}}$, with $\bar{d}$ as a constant, where this form was motivated by experiments~\citep{rubin2011large}. For example, linear elasticity is retrieved for $n=0$ and $F(\phi)=1$ ($\bar{d}=0$). Furthermore, if $F(\phi)=\phi^{\bar{d}}\simeq \left(\frac{\rho}{\rho_s^*}\right)^{\bar{d}}$ is assumed, where $\rho_s^*$ is the unstressed solid density, then depending on the choice of $n$ and $\bar{d}$ the model can recover a variety nonlinear elastic models, such as the ones described by~\citet{alaei2021hydrodynamic,riley2023constitutive,chen2023hydrodynamic}.
% , where $\rho$ is the bulk density and $\rho_s^*$ is the unstressed solid density. Thus, for $n=1$ and $\bar{d}=1$, the model recovers the nonlinear elastic model from~\cref{cha:chap4}.
This versatile structure permits a comprehensive formulation that spans from linear elasticity to pressure-dependent nonlinear elastic material, aptly capturing the behaviour typically observed in experiments.

The perturbed stress of~\cref{eq: hyper_p,eq: hyper_dev} can be determined by first-order Taylor series expansion about the homogeneous unperturbed (reference) strain, which gives
\begin{equation} \label{eq: hyper_pdot}
    \begin{aligned}
        \tilde{p}&=F(\phi)\left[\left(\bar{K}\varepsilon_v^n+\frac{3}{2}n(n-1)\bar{G}\varepsilon_v^{n-2}\varepsilon_s^2\right)\tilde{\varepsilon}_v+3n\bar{G}\varepsilon_v^{n-1}\varepsilon_s\tilde{\varepsilon}_s\right] \\
        &+\frac{ p}{F(\phi)} \frac{\partial F(\phi)}{\partial \phi} \phi\tilde{\varepsilon}_v,
    \end{aligned}
\end{equation}
\begin{equation} \label{eq: hyper_devdot}
    \begin{aligned}
    \tilde{\sigma}_{\alpha\beta}^{\prime}&= F(\phi)\left( 2n\bar{G} \varepsilon_{v}^{n-1} \varepsilon_{\alpha\beta}^{\prime}\tilde\varepsilon_v+ 2\bar{G}\varepsilon_v^{n}\tilde{\varepsilon}_{\alpha\beta}^{\prime}\right) \\
    &+\frac{ \sigma_{\alpha\beta}^{\prime}}{F(\phi)} \frac{\partial F(\phi)}{\partial \phi}\phi \tilde{\varepsilon}_v,
    \end{aligned}
\end{equation}
where $\Box^{*}$ has been dropped from the homogeneous (reference) terms for brevity and the mass balance definition $\tilde{\phi} = \phi \tilde{\varepsilon}_v$ was used, assuming no solid density change, as expected in elastic wave experiments.

Consequently, by substituting perturbed strains~\cref{eq: long_pertub} and the perturbed stress~\cref{eq: hyper_pdot,eq: hyper_devdot} into~\cref{eq: perturbed_mom}, and using the definition $\tilde{\varepsilon}_s=\frac{2}{3}\frac{\varepsilon_{\gamma\zeta}^{\prime}\tilde{\varepsilon}_{\gamma\zeta}}{\varepsilon_s}$, the momentum balance for the hyperelastic model with a plane wave perturbation propagating along $z$ can be written as
% \begin{align} 
% -\rho^{*} \omega^2 &= - k^2 F(\phi) \left(\bar{M}\varepsilon_v^{\frac{b}{1-b}} + \frac{3b(2b-1)}{2(b-1)^2} \bar{G} \varepsilon_v^{\frac{3b-2}{1-b}} \varepsilon_s^2 +\frac{2b}{1-b} \bar{G}\varepsilon_{zz}^{\prime} \varepsilon_v^{\frac{2b-1}{1-b}} \right)\nonumber \\
% &\quad - k^2 \frac{\phi \sigma_{zz}}{F(\phi)} \frac{\partial F(\phi)}{\partial \phi},
% \end{align}
\begin{align} 
&-\rho \omega^2 = - k^2 F(\phi) \left(\bar{M}\varepsilon_v^{n} + \frac{3}{2} n(n-1)\bar{G} \varepsilon_v^{n-2} \varepsilon_s^2 \right. \nonumber \\
& \left.+2n\bar{G}\varepsilon_v^{n-1}\varepsilon_{zz}^{\prime} +2n \bar{G}\varepsilon_v^{n-1}\sum_{\beta=x}^{z}\varepsilon_{z\beta}^{\prime}  \right) - k^2 \frac{\phi \sigma_{zz}}{F(\phi)} \frac{\partial F(\phi)}{\partial \phi},
\end{align}
where $\bar{M}=\bar{K}+\frac{4}{3}\bar{G}$ is the constrained modulus, and all state variables are defined in terms of homogeneous unperturbed states. Notably, this result showcases that while the perturbation is along the z-axis, there emerges a dependence on the unperturbed states in the state of other directions, \emph{i.e.}, $\varepsilon_{z\beta}^{\prime}$. The wave speed of longitudinal waves is defined as $V_p=\frac{\omega}{k}$~\citep{landau1986theory,achenbach2012wave}, which gives
% \begin{align} \label{eq: vp_general_hyper}
%     V_{p} & = \frac{\omega}{k} \notag \\
%     &=\sqrt{\frac{ F(\phi) \left(\bar{M}\varepsilon_v^{\frac{b}{1-b}} + \frac{3b(2b-1)}{2(b-1)^2} \bar{G} \varepsilon_v^{\frac{3b-2}{1-b}} \varepsilon_s^2 +\frac{2b}{1-b} \bar{G}\varepsilon_{zz}^{\prime} \varepsilon_v^{\frac{2b-1}{1-b}} \right) + \frac{\phi \sigma_{zz}}{F(\phi)} \frac{\partial F(\phi)}{\partial \phi}   }{\rho^*}}.
%  \end{align}
\begin{align} \label{eq: vp_general_hyper}
    & \rho\left(V_{p}^{hyper}\right)^2= F(\phi) \left(\bar{M}\varepsilon_v^{n} + \frac{3}{2} n(n-1)\bar{G} \varepsilon_v^{n-2} \varepsilon_s^2 \right. \notag \\
    &\left.+2n\bar{G}\varepsilon_v^{n-1}(\varepsilon_{zx}^{\prime} + \varepsilon_{zy}^{\prime} + 2\varepsilon_{zz}^{\prime}) \right) + \frac{\phi \sigma_{zz}}{F(\phi)} \frac{\partial F(\phi)}{\partial \phi} .
\end{align}

Similarly, by substitution of~\cref{eq: shear_perturb} and the perturbed stress~\cref{eq: hyper_pdot,eq: hyper_devdot} into~\cref{eq: perturbed_mom}, the transverse wave speed of a wave propagating along $z$-direction and polarised along $x$-direction can be found to be
\begin{equation} \label{eq: vs_general_hyper}
     \rho\left(V_{s}^{hyper}\right)^2 = F(\phi)\bar{G} \varepsilon_v^{n-1}\left(n\varepsilon_{zx}^{\prime}+\varepsilon_v\right).
\end{equation}

 Notably, the longitudinal wave speed in~\cref{eq: vp_general_hyper} and transverse wave speed in~\cref{eq: vs_general_hyper} both depend on strain (or stress states) that cannot be represented solely by invariants, which are a result of the chosen wave perturbation direction and model. Thus, if different propagation or polarisation directions other than the ones illustrated in~\cref{fig: 1c5} were used on the same unperturbed state, the wave speed would naturally depend on different strain components. However, as shown later, only the longitudinal wave speed maintains this dependence on directionality for triaxial conditions.

\subsection{Tangent moduli for hyperelastic model} \label{section: t_moduli}
The particular choice of the internal energy density~\cref{eq: internal energy} is motivated here. To this end, we will show that the tangent (instantaneous) moduli for the simplified scenario of zero shear strain is equivalent to~\cref{eq: M_iso_emp,eq: G_iso_emp}. To begin, the  constrained tangent modulus  is determined by $M_t=C_{zzzz}=\frac{\partial \sigma_{zz}}{\partial \varepsilon_{zz}}$, which is found to depend on the volumetric strain and solid fraction as follows:
% \begin{equation}
%     \begin{aligned}
%     M_t &=\frac{\partial \sigma_{zz}}{\partial \varepsilon_{zz}} \\
%     &= F(\phi)\left(\bar{M}\varepsilon_v^n+\frac{3}{2}n(n-1)\bar{G}\varepsilon_v^{n-2}\varepsilon_s^2+2n\bar{G}\varepsilon_v^{n-1}\varepsilon_s + 2n\bar{G} \varepsilon_{v}^{n-1} \varepsilon_{zz}^{\prime}\right) + \frac{\phi \sigma_{zz}}{F(\phi)} \frac{\partial F(\phi)}{\partial \phi}.
%     \end{aligned}
% \end{equation}

\begin{equation}
    \begin{aligned}
    M_t &= F(\phi)\left(\bar{M}\varepsilon_v^n+\frac{3}{2}n(n-1)\bar{G}\varepsilon_v^{n-2}\varepsilon_s^2\right. \\
    &\left.+ 4n\bar{G}\varepsilon_v^{n-1}\varepsilon_{zz}^{\prime}\right) + \left(\frac{\phi}{F(\phi)} \frac{\partial F(\phi)}{\partial \phi}\right)\sigma_{zz}.
        \end{aligned}
\end{equation}
Importantly, the right-hand side of the longitudinal wave speed in~\cref{eq: vp_general_hyper} is equivalent to this modulus in the absence of deviatoric shear components $\varepsilon_{xz}^{\prime}=\varepsilon_{yz}^{\prime}=0$. 
% \begin{equation} \label{eq: instant_G_hyper}
%     G_t = \frac{\partial q}{\partial \varepsilon_s} = 3F(\phi)\bar{G}\varepsilon_v^n.
% \end{equation}
% where $\bar{M}=\bar{K}+\frac{4}{3}\bar{G}$. 

In the absence of shear strain, which happens under isotropic stress states whereby the shear stresses are zero, it is readily shown by substituting~\cref{eq: hyper_p} that the constrained modulus simplifies to
% \begin{equation} \label{eq: instant_M_iso}
%     , \quad b=\frac{n}{1+n},
% \end{equation}
\begin{subequations}
\begin{align}
M_t& =  F(\phi) \bar{M}\varepsilon_v^{\frac{b}{1-b}}+ \left(\frac{\phi }{F(\phi)} \frac{\partial F(\phi)}{\partial \phi}\right)p, \label{eq: insta_M_isoa} \\
b&=\frac{n}{1+n}, \label{eq: insta_M_isob}
\end{align}
\end{subequations}
% \begin{equation} \label{eq: instant_M_isoa} 
%     M_t= F(\phi) \bar{M}\varepsilon_v^{\frac{b}{1-b}}+ \left(\frac{\phi }{F(\phi)} \frac{\partial F(\phi)}{\partial \phi}\right)p
% \end{equation}
where we note an additional linear dependence on pressure, a structure that seems to differ greatly from observed empirical trends.

% However,~\cref{eq: insta_M_isoa} in conjunction with~\cref{eq: hyper_p} and substitution of  $F(\phi)=\phi^{\bar{d}}$ can be rewritten in the following form
% \begin{equation}\label{eq: insta_M_iso1.5}
%      M_t =  \left( \frac{p}{(1-b) \varepsilon_v} + \frac{4}{3}\bar{G}\phi^{\bar{d}}\varepsilon_v^{\frac{b}{1-b}} \right) + \bar{d}p.
% \end{equation}
% Assuming small elastic volumetric deformations $\varepsilon_v \ll 1$, the first term on the right-hand side is larger than the density correction provided by the last term. For a reasonable choice of $F(\phi)=\phi^{\bar{d}}$ for the density dependence, then it is clear that
% \begin{equation}
%     F(\phi)\bar{M}\varepsilon^{\frac{b}{1-b}}\gg\left(\frac{\phi}{F(\phi)} \frac{\partial F(\phi)}{\partial \phi}\right)p.
%     \label{eq:negligible}
% \end{equation}

However, by recognising that $p\propto F(\phi) \bar{M}\varepsilon_v^{\frac{1}{1-b}}$, it is apparent that $F(\phi) \bar{M}\varepsilon_v^{\frac{b}{1-b}} \propto \frac{p}{\varepsilon_v}$. Furthermore, considering the reasonable and general choice~\citep{rubin2011large} of $F(\phi)=\phi^{\bar{d}}$ the second term reduces to $\bar{d}p$. It is now clear that when assuming small elastic volumetric deformations $\varepsilon_v \ll 1$ that
\begin{equation}
    F(\phi)\bar{M}\varepsilon^{\frac{b}{1-b}}\gg\left(\frac{\phi}{F(\phi)} \frac{\partial F(\phi)}{\partial \phi}\right)p.
    \label{eq:negligible}
\end{equation}
Therefore, the approximate constrained tangent modulus can be expressed through substitution of~\cref{eq: hyper_p} 
\begin{subequations}
\begin{align}
M_t & =\frac{\partial \sigma_{zz}}{\partial \varepsilon_{zz}} \approx A H(\phi) \left(\frac{p}{p_a}\right)^b, \label{eq: instant_M_iso2a} \\
A & = \bar{M}\left( \frac{p_a}{(1-b)\bar{K}}\right)^{b}, \label{eq: instant_M_iso2b} \\
&H(\phi)=F(\phi)^{1-b},
\end{align}
\end{subequations}
where $p_a$ was introduced such that $A$ has units of stress. Similarly, the shear tangent modulus $G_t$ with respect to triaxial shear strain with substitution of~\cref{eq: hyper_p} is
\begin{subequations}
\begin{align}
G_t & =\frac{1}{3}\frac{\partial q}{\partial \varepsilon_s}=  B H(\phi)  \left(\frac{p}{p_a}\right)^b, \label{eq: instant_G_isoa} \\
B & = \bar{G}\left( \frac{p_a}{(1-b)\bar{K}} \right)^{b}, \label{eq: instant_G_isob}
\end{align}
\end{subequations}
% \begin{equation} \label{eq: instant_G_iso}
%     G_t = \frac{\partial q}{\partial \varepsilon_s}=  B H(\phi)  \left(\frac{p}{p_a}\right)^b, \quad  B=\bar{G}\left( \frac{p_a}{(1-b)\bar{K}} \right)^{b},
% \end{equation}
where again $p_a$ was introduced such that $B$ has units of stress. Thus, under the assumption of isotropic conditions and small elastic strains ($\varepsilon_s=0$, $\varepsilon_v\ll 1$), the tangent moduli reduce to the experimental forms found previously~\cref{eq: M_iso_emp,eq: G_iso_emp} in the hyperelastic model. 

\subsection{Hypoelastic wave speed} \label{section: hypo_waves}
Hypoelastic models tend to adopt the experimentally developed empirical stiffness in~\cref{eq: M_iso_emp,eq: G_iso_emp} without the requirement of being determined from internal energy. Thus, hypoelastic models exhibit path-dependency and do not conform to the thermodynamic requirements of energy conservation~\citep{zytynski1978modelling,einav2004pressure}. The stress rates determined directly from the empirical relations (\cref{eq: M_iso_emp,eq: G_iso_emp}) are
\begin{equation} \label{eq: hypo_pdot}
        \dot{p}=\bar{K}^{hypo}H(\phi)\left(\frac{p}{p_a}\right)^b\dot{\varepsilon}_v,
\end{equation}
\begin{equation} \label{eq: hypo_devdot}
    \dot{\sigma}_{\alpha\beta}^{\prime}= 2\bar{G}^{hypo} H(\phi)\left(\frac{p}{p_a}\right)^b\dot{\varepsilon}_{\alpha\beta}^{\prime},
\end{equation}
where $\bar{K}^{hypo}$ and $\bar{G}^{hypo}$ are the bulk and shear stiffness constants for the hypoelastic model, which have a stress dimension. The superscript emphasises that although these are elastic constants, their values is not inherently equal to those of the hyperelastic model.

Following a procedure similar to the derivation of the hyperelastic wave speeds, the perturbed stress is substituted into~\cref{eq: perturbed_mom}. However, hypoelasticity does not have an explicit expression for total stress $\sigma_{\alpha\beta}$ and as such, it assumed that $\tilde{\sigma}_{\alpha\beta}=\dot{\sigma}_{\alpha\beta}$ as this was the result of the first-order Taylor expansion of~\cref{eq: hyper_p,eq: hyper_dev}. The resulting longitudinal and transverse wave velocities are
% \begin{equation} \label{eq: vp_general_hypo}
%     V_p=\sqrt{\frac{ A  H(\phi) \left(\frac{p}{p_a}\right)^{b}    }{\rho^*}}, \quad A = \left(\bar{K}^{hypo}+\frac{4}{3}\bar{G}^{hypo}\right)
% \end{equation}
% \begin{equation}\label{eq: vs_general_hypo}
%      V_{s} = \sqrt{\frac{B  H(\phi) \left(\frac{p}{p_a}\right)^{b} }{\rho^*}},  \quad B= 2\bar{G}^{hypo}
% \end{equation}
\begin{subequations}
\begin{align}
V_{p}^{hypo} & = \sqrt{\frac{ A  H(\phi)}{\rho} \left(\frac{p}{p_a}\right)^{b}    },  \label{eq: vp_general_hypoa} \\
A & = \left(\bar{K}^{hypo}+\frac{4}{3}\bar{G}^{hypo}\right),  \label{eq: vp_general_hypob}
\end{align}
\end{subequations}
\begin{subequations}
\begin{align}
V_{s}^{hypo} & = \sqrt{\frac{B  H(\phi)}{\rho} \left(\frac{p}{p_a}\right)^{b} },  \label{eq: vs_general_hypoa} \\
B & =  \bar{G}^{hypo},  \label{eq: vs_general_hypob}
\end{align}
\end{subequations}
which are equivalent to~\cref{eq: vp_elastic,eq: vs_elastic} when~\cref{eq: M_iso_emp,eq: G_iso_emp} are substituted, and thus, the empirical relations are recovered. However, the hypoelastic wave speeds lack the additional stress-dependent terms that emerged in the hyperelastic derivation due to the energy conservation of the latter model. These differences are critical to the comparison against experimental results that will be shown between the hyperelastic and the hypoelastic model in triaxial compression. Notably, $A$ and $B$ in~\cref{eq: vp_general_hypob,eq: vs_general_hypob} depend on the elastic constants differently than their counterparts in~\cref{eq: instant_M_iso2b,eq: instant_G_isob}.

\subsection{Linear elastic limit}
For both the hyperelastic and hypoelastic models, linear elasticity is recovered when $b=0$ and $H(\phi)=1$. In this limit, the longitudinal wave speed and transverse wave speed for both hyperelastic and hypoelastic models reduce to
\begin{equation} 
    V_p=\sqrt{\frac{\bar{M}}{\rho}},
\end{equation}
\begin{equation} 
    V_s=\sqrt{\frac{\bar{G}}{\rho}},
\end{equation}
where the superscript notation is dropped as the stiffness constants for the hyperelastic and hypoelastic models are equivalent in this case. Therefore,~\cref{eq: vp_elastic,eq: vs_elastic} are recovered in the linear elastic limit.

\section{Comparison with experimental results}

\subsection{Isotropic compression}
In the isotropic compression scenario (\emph{i.e.}, $\varepsilon_s=\varepsilon_{\alpha\beta}^{\prime}=0$), we show that the hyperelastic wave speeds~\cref{eq: vp_general_hyper,eq: vs_general_hyper} reduce to the same form as the wave speeds of the hypoelastic model~\cref{eq: vp_general_hypoa,eq: vs_general_hypoa}. If~\cref{eq: hyper_p} while $\varepsilon_s=0$ (isotropic compression) is substituted into~\cref{eq: vp_general_hyper}, then the longitudinal wave speed is
\begin{equation}\label{eq: vpiso_approx}
        V_{p}^{hyper} \approx \sqrt{\frac{ A  H(\phi)}{\rho} \left(\frac{p}{p_a}\right)^{b} }.
\end{equation}
where $A$ is defined in~\cref{eq: instant_M_iso2b} and the density correction term, \emph{i.e.},  $\frac{\phi p}{F(\phi)} \frac{\partial F(\phi)}{\partial \phi} $, is ignored as discussed previously (cf. Eq.~\ref{eq:negligible}). Similarly, by substitution of~\cref{eq: hyper_p} into~\cref{eq: vs_general_hyper}, the corresponding transverse wave speed for the hyperelastic model is
\begin{equation} \label{eq: vs_iso_analytical}
     V_{s}^{hyper}=\sqrt{\frac{B H(\phi)}{\rho} \left(\frac{p}{p_a}\right)^{b}},
\end{equation}
 where $B$ is defined in~\cref{eq: instant_G_isob}. The derived wave speeds are equivalent to the hypoelastic model (\cref{eq: vp_general_hypoa,eq: vs_general_hypoa}). However, it is crucial to note that $A$ and $B$ are combined parameters of the bulk and shear stiffness constants, $\bar{K}$ and $\bar{G}$, respectively. Therefore, to accurately determine the stiffness constants, both $V_p$ and $V_s$ must be measured for the hyperelastic and hypoelastic models.
\begin{table}[t!]
\caption{Summary of effective model constants obtained for fitting with the isotropic experimental measurements. Note that $d=(1-b)\bar{d}$ for the hyperelastic model.}
\centering
\begin{tabular}{|l|l|l|l|}
\hline
 $A$ (MPa) & $B$ (MPa) & $b$ & $d$   \\
\hline
95.27 & 31.40 & 0.35 & 5.32\\
\hline
\end{tabular}
\label{table: model fits}
\end{table}

\begin{table}[t!]
\caption{Summary of model stiffness constants obtained for fitting with the isotropic experimental measurements. The stiffness constants $\bar{K}$ and $\bar{G}$ are determined from~\cref{eq: instant_M_iso2b,eq: instant_G_isob}, while the stiffness constants $\bar{K}^{hypo}$ and $\bar{G}^{hypo}$ are determined from~\cref{eq: vp_general_hypob,eq: vs_general_hypob}.}
\centering
\centering
\begin{tabular}{|l|l|l|l|}
\hline
 $\bar{K}$ (MPa) & $\bar{G}$ (MPa) & $\bar{K}^{hypo}$ (MPa) & $\bar{G}^{hypo}$ (MPa)  \\
\hline
62.1$\cdot 10^{4}$&36.5$\cdot 10^{4} $& 40.04& 31.40\\
\hline
\end{tabular}
\label{table: model stiffness fits}
\end{table}

\begin{figure}[!htbp]
	\centering
	\includegraphics[scale=0.4]{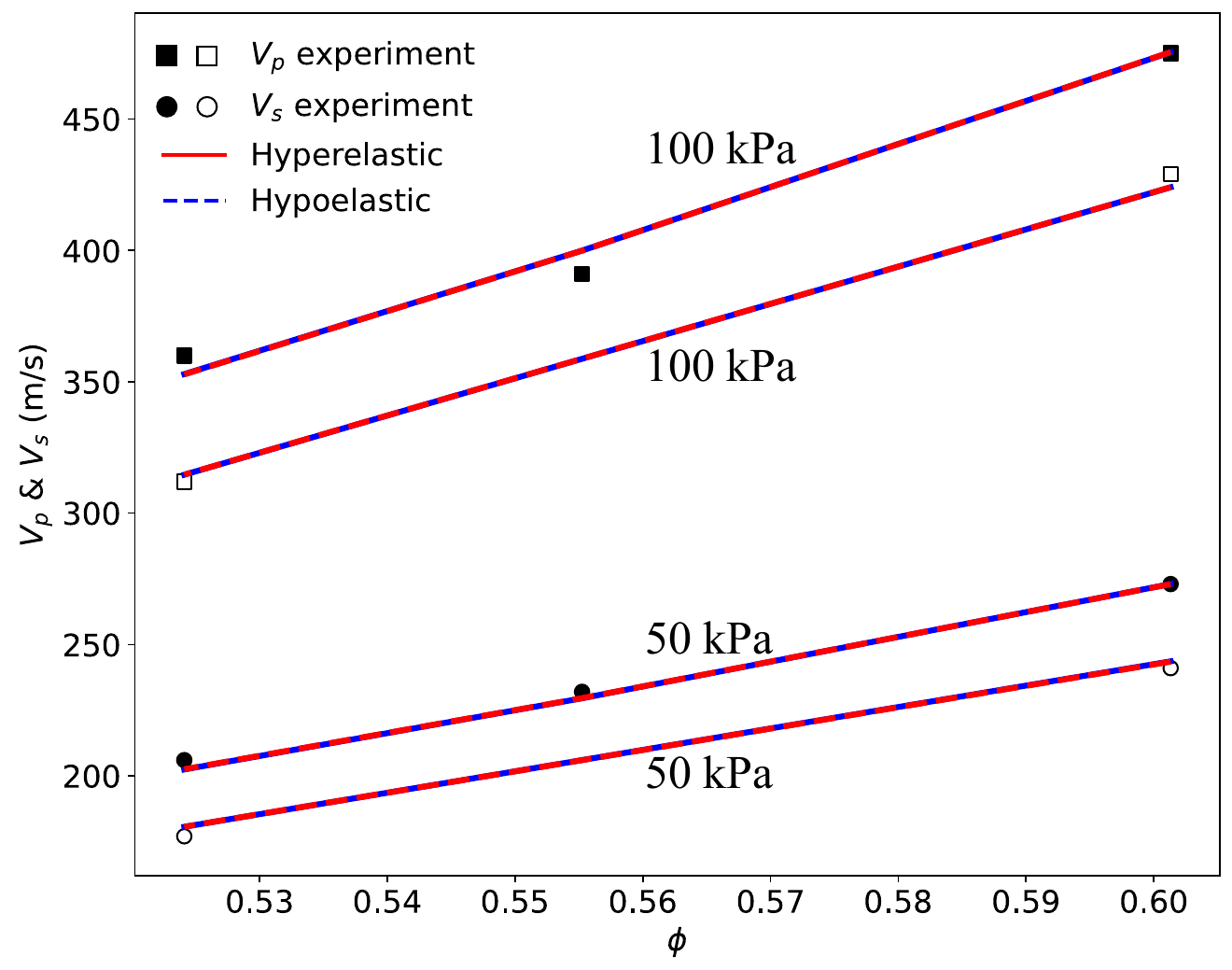}
  	\caption{Transverse wave velocity $V_s$ and longitudinal wave velocity $V_p$ against solid fraction $\phi$ for an isotropic stress state of 50 and 100 kPa. The empirical relations are not shown because they are proven to be equivalent to the wave speeds derived from the hypoelastic model. The experimental data was extracted from~\cite{dutta2021effect}.}
   % ; see~\cref{section: hypo_waves}.}
	\label{fig: 2c5}       % Give a unique label
\end{figure}
Both models are fit to an experimental dataset of wave velocities to elucidate better the equivalence of these derived wave speeds and the empirical relations~\cref{eq: M_iso_emp,eq: G_iso_emp}. The data was acquired by digitising results presented in~\cite{dutta2021effect}. The wave measurement was performed on Toyoura sand with a specific gravity of $G_s=2.64$, a minimum void ratio of $e_{min}=0.569$, and a maximum void ratio of $e_{max}=0.936$. Notably, $V_p$ and $V_s$ were fit to experimental measurements simultaneously. Despite the sparsity of data, both models fit the experimental results well, and their equivalency in this scenario is clear. The resulting coefficients are shown in~\cref{table: model fits}, which will also be used for the following section.  The stiffness constants for the hyperelastic and the hypoelastic model are shown in~\cref{table: model stiffness fits}. Notably, the models have several order of magnitude differences between the elastic coefficients, but it is important to recall that these are from two different models, and therefore cannot be directly compared together. Moreover, the rather large elastic stiffness values of the hyperelastic model occur because once~\cref{eq: instant_M_iso2b,eq: instant_G_isob} are substituted into~\cref{eq: vpiso_approx,eq: vs_iso_analytical} it becomes clear that the pressure is scaled by $\bar{K}$ as opposed to $p_a$ as in the hypoelastic model.

% Thus, stiffness constants $\bar{K}$ and $\bar{G}$ are determined from~\cref{eq: instant_M_iso2b,eq: instant_G_isob}, while the stiffness constants $\bar{K}^{hypo}$ and $\bar{G}^{hypo}$ are determined from~\cref{eq: vp_general_hypob,eq: vs_general_hypob}. Despite the sparsity of data, both models fit the data sets well, and their equivalency in this scenario is clear. The resulting coefficients are shown in~\cref{table: model fits}, which will also be used for the following section. 

\subsection{Triaxial compression} \label{sect: triax}
While the wave speeds of both models are shown to be mathematically equivalent for isotropic compression, this is not the case for triaxial states. For the hypoelastic model, wave speeds do not change from~\cref{eq: vp_general_hypoa,eq: vs_general_hypoa} because they solely depend on $\phi$ and $p$. However, the hyperelastic model shows that stress dependence is more involved in triaxial states due to being derived from an energy potential.~\cref{fig: 2c5} (b) shows the wave perturbations and relevant directions of propagation and particle motion along principle directions in this scenario. For the hyperelastic model, the longitudinal wave speed in triaxial conditions is equivalent to~\cref{eq: vp_general_hyper}, but since the reference state is in triaxial conditions, $\varepsilon_{zx}^{\prime}=\varepsilon_{zy}^{\prime}=0$ and only $\varepsilon_{zz}^{\prime} \neq 0$,
while the transverse wave speed~\cref{eq: vs_general_hyper}, by substitution of~\cref{eq: hyper_p}, becomes 
% \begin{equation}\label{eq: vp_triax_hyper}
%         V_{p}^{hyper} =\sqrt{\frac{ F(\phi)  \left(\bar{M}\varepsilon_v^{\frac{b}{1-b}} + \frac{3b(2b-1)}{2(b-1)^2} \bar{G} \varepsilon_v^{\frac{3b-2}{1-b}} \varepsilon_s^2 +\frac{2b}{1-b} \bar{G}\varepsilon_{zz}^{\prime} \varepsilon_v^{\frac{2b-1}{1-b}} \right) }{\rho}}.
% \end{equation}
\begin{equation} \label{eq: vs_triax_hyper}
    V_{s}^{hyper}=\sqrt{\frac{\bar{G}H(\phi)}{\rho} \left( \frac{p}{2(1-b)\bar{K}}\left( 1 + \sqrt{1 - \frac{2b\bar{K}\eta^2}{3\bar{G}}} \right) \right)^{b}}
\end{equation}
where $\eta=\frac{q}{p}$, and $q=\sqrt{\frac{3}{2}\sigma_{\alpha \beta}^{\prime}\sigma_{\alpha \beta}^{\prime}}$ is the triaxial shear stress. Notably, both $V_p^{hyper}$ and $V_s^{hyper}$ can become imaginary when $\eta=\sqrt{\frac{3 \bar{G}}{2b\bar{K}}}$, which indicates material instability~\citep{rice1976localization,zhang2012elastic,stefanou2019strain}. Such instabilities have also been documented in other hyperelastic models by evaluating the convexity of the energy~\citep{jiang2003granular,agnolin2007internal}. Moreover, the transverse wave speed depends only on the stress invariants; thus, for triaxial states, the wave perturbation direction plays no role in the wave speed when the wave propagates along a principal direction as illustrated in~\cref{fig: 2c5}.

\begin{figure*}[t]
	\centering
	\includegraphics[width=\linewidth]{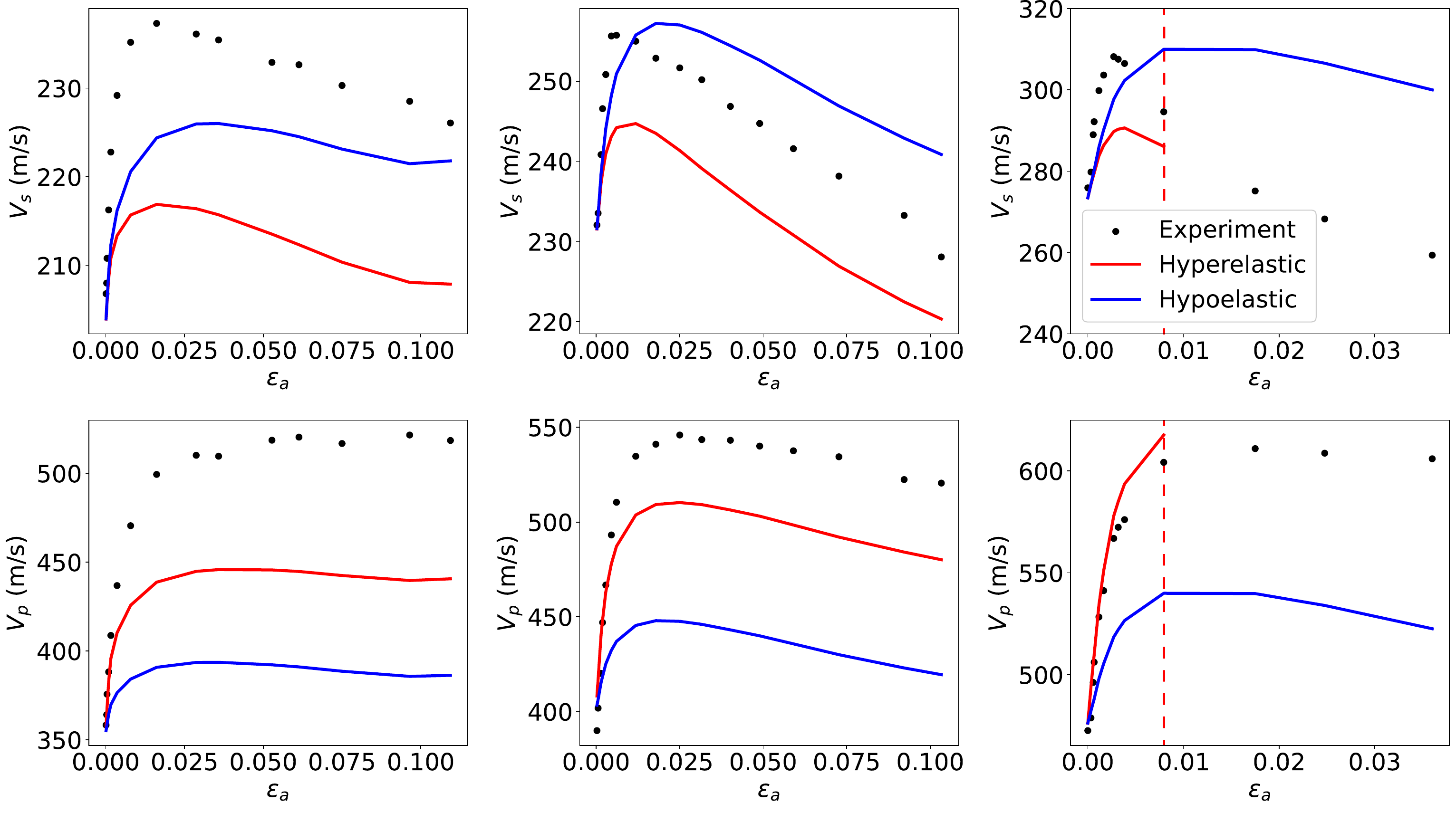}
   	\caption{Variation of $V_s$ (top row) and $V_p$  (bottom row) against axial strain $\varepsilon^a$ for (a) $\phi_0\approx 0.52$, (b) $\phi_0\approx 0.55$, and (c) $\phi_0\approx 0.60$. In (c), the red dashed line highlights the point beyond which the hyperelastic model predicts material instability. Additionally, the experimental results in (c) are limited to results up to the peak stress that occurred at $\varepsilon^a \approx0.04$ as the experiment underwent a shear band, as stated in~\cite{dutta2021effect}. The experimental data was extracted from~\cite{dutta2021effect}.}
	\label{fig: 3c5}       % Give a unique label
\end{figure*}

To understand the differences between the models, wave speed predictions are compared with wave measurements obtained from experimental data, which were digitised from results presented by~\cite{dutta2021effect}. The triaxial experiment conducted was a drained compression test on dry sand, ensuring that the wave speeds would not be influenced by the presence of water~\citep{cho2001unsaturated,barriere2012laboratory,leong2016effects}. The experiments were performed under a radial stress $\sigma_{r}=100$ kPa for three different initial solid fractions: $\phi_0 \approx 0.52$, $\phi_0 \approx 0.55$, and $\phi_0 \approx 0.60$. It is crucial to highlight that the densest experiment ($\phi_0 \approx 0.60$) experienced shear localisation~\citep{dutta2021effect}. Therefore, experimental results are shown only up to the peak stress ($\varepsilon^a\approx0.04$) for this experiment, and specimen homogeneity cannot be assumed. There are no free parameters in the prediction at this stage since the stiffnesses obtained from the previous section in isotropic experiments are kept the same, which are shown in~\cref{table: model stiffness fits}.

% Stiffness parameters were determined from $A$ and $B$ with~\cref{eq: instant_M_iso2b,eq: instant_G_isob} for the hyperelastic model, and with~\cref{eq: vp_general_hypob,eq: vs_general_hypob} for the hypoelastic model. 

\cref{fig: 3c5} plots the transverse and longitudinal velocities against the axial strain. The hyperelastic model under-predicts the experimental data for the transverse wave velocities, whereas the hypoelastic model under or over-predicts experimental data depending on the initial density. However, the hyperelastic model generally offers a more accurate prediction to the data for longitudinal wave speed than the hypoelastic model. Notably, for $\phi_0\approx 0.60$, the hyperelastic model could not produce a prediction beyond the red dotted line in ~\cref{fig: 3c5} (c) as these stress states resulted in an imaginary wave speed, suggesting the emergence of an instability. However, this occurred earlier than the peak stress of the experiment.

\begin{figure*}[t]
	\centering
	\includegraphics[width=\linewidth]{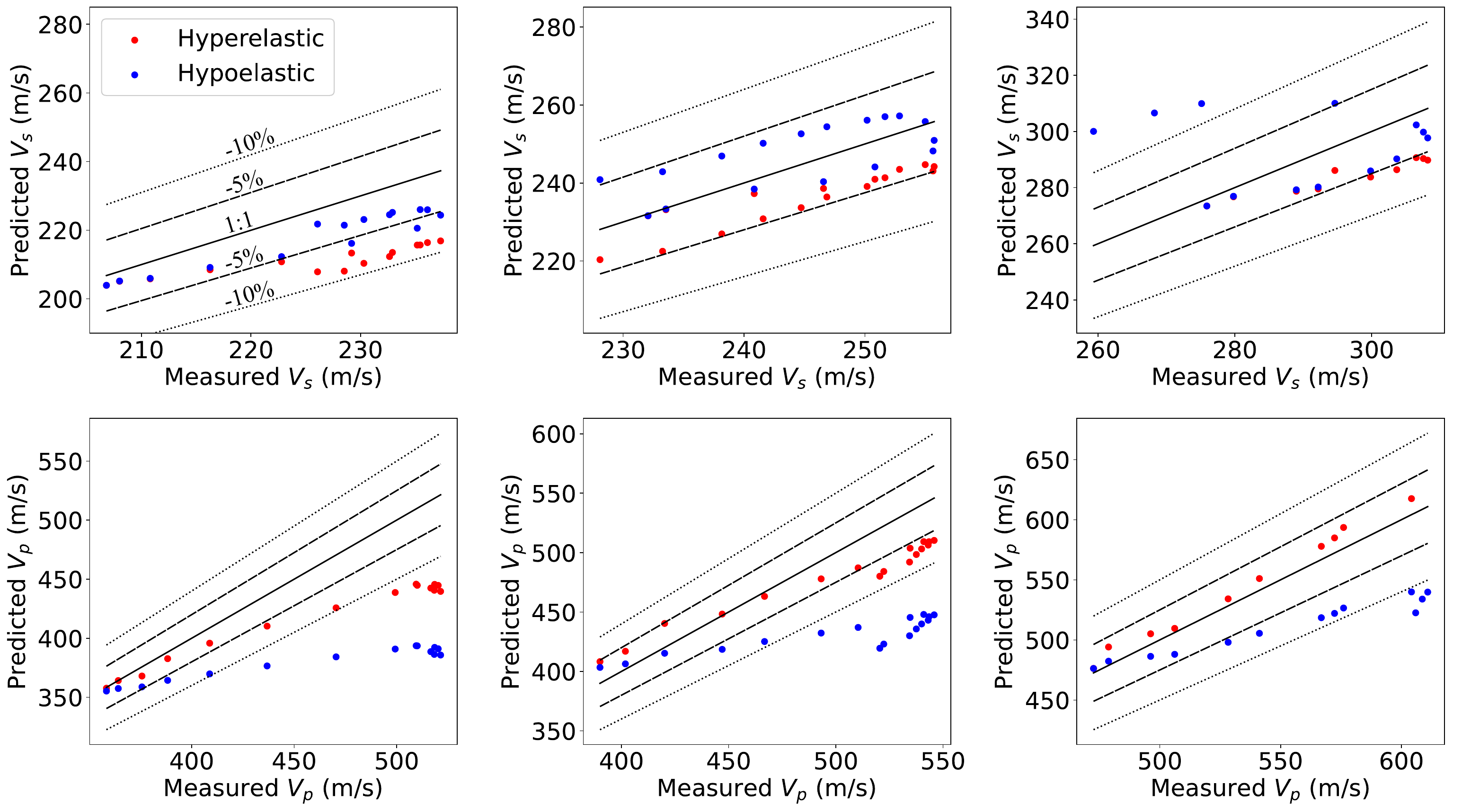}
   	\caption{Predicted $V_s$ (top row) and $V_p$  (bottom row) against experimentally measured velocities for (a) $\phi_0\approx 0.52$, (b) $\phi_0\approx 0.55$, and (c) $\phi_0\approx 0.60$. The solid black line denotes the agreement line with a slope of 1, the dashed line represents $\pm 5$\% error of a perfect prediction, and the dotted line is $\pm 10$\% error.}
	\label{fig: 4c5}       % Give a unique label
\end{figure*}

The observations are further substantiated by~\cref{fig: 4c5}, which compares the measured values with those predicted by both models. Encouragingly, both models predicted wave speeds are within a range of $\pm$10\% of the measured values, although the hyperelastic model tends to yield better results overall. Both models accurately reproduce shear wave velocities, although the hyperelastic model exceeds a $-5$\% error for $\phi_0\approx0.52$. However, the hyperelastic model outperforms predicting longitudinal wave speeds and essentially stays within the $\pm10$\% error range for $\phi_0\approx 0.55$ and $\phi_0\approx 0.60$. The hypoelastic model exceeds a $-10$\% error in these cases. These results suggest that the additional stress-dependent terms in the hyperelastic model provide correction factors missing in the hypoelastic models.

\begin{figure}[t!]
	\centering
	\includegraphics[width=\linewidth]{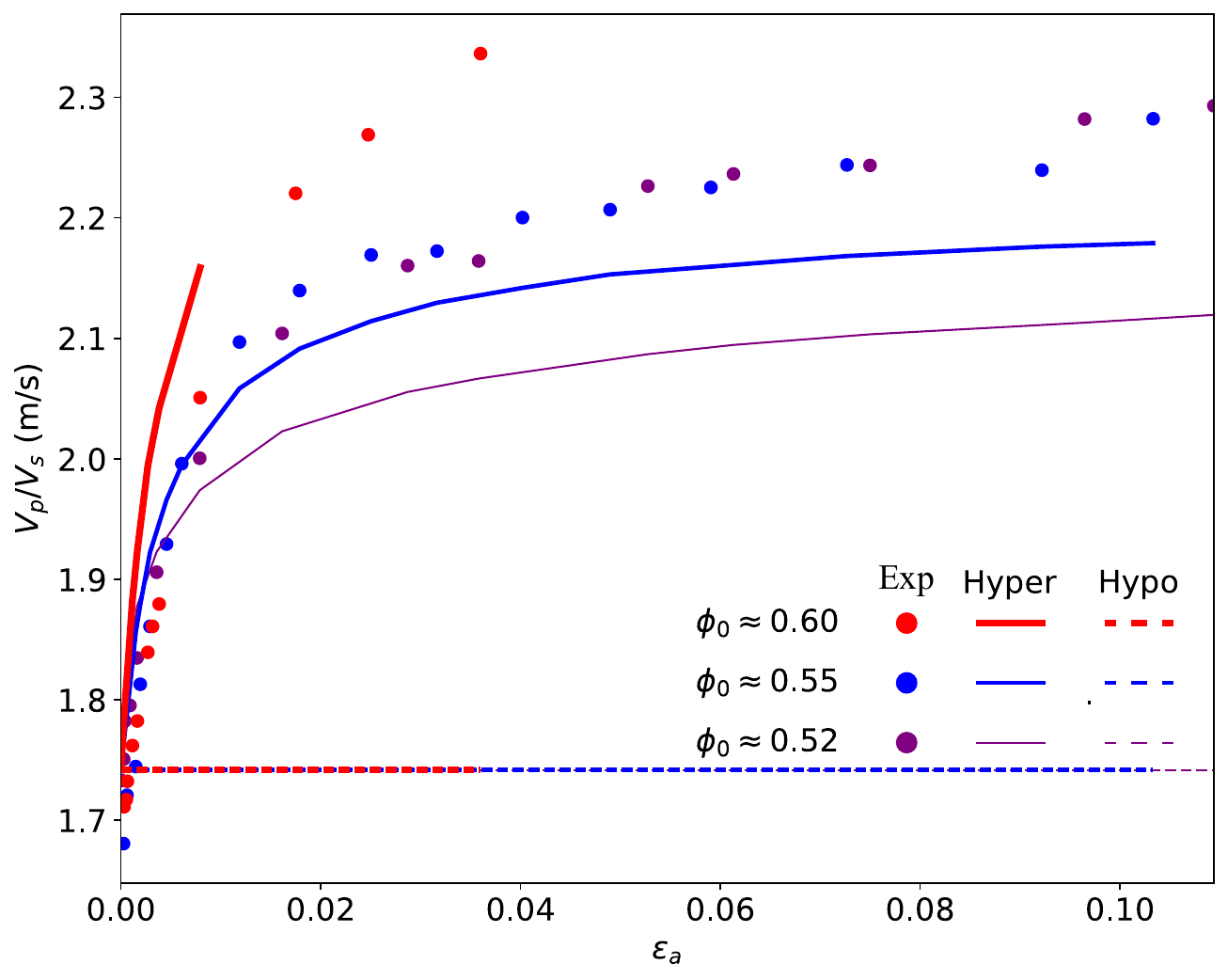}
	\caption{Variation of $V_p/V_s$ against axial strain $\varepsilon^a$. The results for the experiment (solid points), the hyperelastic model (solid line), and the hypoelastic model (dashed line) are plotted for $\phi\approx0.60 $ (red), $\phi\approx0.55 $ (blue). }
	\label{fig: 5c5}       % Give a unique label
\end{figure}
\subsection{Material isotropy}
Lastly, the $V_p/V_s$ ratio is examined to explore shearing-induced differences in the evolution of $V_p$ and $V_s$.~\cref{fig: 5c5} illustrates the variation of $V_p/V_s$ against $\varepsilon^a$. Notably, the data reveals an initial increase of $V_p/V_s$ before plateauing, with $\phi_0\approx 0.60$ not exhibiting a plateau as the experimental results are not shown beyond $\varepsilon^a\approx0.04$ (peak stress). However, the hypoelastic model maintains a constant ratio because the power law coefficients are assumed to be identical for both $V_p$ and $V_s$ as is required for isotropic materials. One might argue that the power coefficients should differ for empirical relations~\cref{eq: M_iso_emp,eq: G_iso_emp} or include fabric dependency~\citep{dutta2020evolution,li2021analysis}, leading to some evolution of $V_p/V_s$, and therefore, extending the empirical models to anisotropic materials. However, the hyperelastic model captures this evolution without additional assumptions (or extra parameters) thanks to the additional stress-dependent terms that arise when deriving the wave speeds. The prediction of evolving $V_p/V_s$ suggests that, at least to a first-order, incorporating a more sophisticated stress dependency or fabric might not be essential as discussed in~\citet{mayer2010propagation}. This offers a compelling methodology to accurately account for stress-induced anisotropy and inherent material anisotropy, enhancing our understanding of prevailing mechanisms within soils.

\section{Conclusions}
 
In this paper, wave speeds were derived for hyperelastic and hypoelastic models that depend on pressure and density in a general manner. The hypoelastic model's wave speed is equivalent to the empirical relations. The hyperelastic model is also equivalent to the empirical relations for isotropic compression.

However, for triaxial states, the wave speeds of the hyperelastic model become more elaborate, revealing previously unobserved stress dependencies, which are not present in the hypoelastic model. The hypoelastic model, equivalent to the empirical relations, generally mirrors the overall trends in longitudinal and transverse wave speed. Yet, it does not accurately capture the evolution of their ratios $V_p/V_s$. Conversely, the hyperelastic model predicts an evolution in the ratios of the longitudinal and transverse waves due to these additional stress dependencies arising from terms ensuring energy conservation. This suggests that material anisotropy may not be required for a first-order approximation of the evolution of $V_p/V_s$. Furthermore, these intricate stress dependencies might illuminate the observed evolution of $V_p$ under high stresses in oedometric compression~\citep{jia1999ultrasound,makse2004granular}. More specifically, these findings emerge from not losing energy while waves travel in hyperelastic media, unlike the case of hypoelastic media, and from the consistently derived wave speeds for pressure-dependent elastic media.

Additionally, it is paramount to highlight that the analytical derivation allows the isolation of the true elastic constants of the material, in contrast to the aggregated parameters frequently assumed in empirical correlations. This distinction is pivotal, as elastic stiffness often serves as a benchmark for design methodologies~\citep{atkinson2000non,clayton2011stiffness} and model calibration~\citep{fahey1993finite,ayala2022computational,huang2023small}.

Finally, the derived transverse wave speed for hyperelasticity indicates that the wave propagation direction does not affect triaxial states when propagating along principal directions. This suggests that there is a necessity for either fabric or an anisotropic constitutive model to reproduce experimental results for materials that produce this effect~\citep{kuwano2002applicability,zamanian2020directional,liu2022anisotropy}. Furthermore, this procedure could be extended to incorporate other state variables, such as fabric, grain size and shape, as well as saturation. Thus, it provides a methodological and consistent manner in which the effect of additional state variables on wave speed can be considered.

\bibliography{apssamp}% Produces the bibliography via BibTeX.

\end{document}